\documentclass{article}
\usepackage{amssymb}
\usepackage{ijcai25}

\usepackage{times}
\usepackage{soul}
\usepackage{url}
\usepackage[hidelinks]{hyperref}
\usepackage[utf8]{inputenc}
\usepackage[small]{caption}
\usepackage{graphicx}
\usepackage{amsmath}
\usepackage{amsthm}
\usepackage{booktabs}
\usepackage{algorithm}
\usepackage{algorithmic}
\usepackage[switch]{lineno}
\usepackage{graphicx} 
\usepackage{multirow}
\usepackage{wrapfig}

\title{Soft Reasoning Paths for Knowledge Graph Completion}

\author{
Yanning Hou$^1$\and
Sihang Zhou$^1$\thanks{Corresponding author.}\and
Ke Liang$^2$\and
Lingyuan Meng$^2$\and
Xiaoshu Chen$^2$\and
Ke Xu$^3$\and
Siwei Wang$^2$\and
Xinwang Liu$^2$\and
Jian Huang$^1$\\
\affiliations
$^1$College of Intelligence Science and Technology, National University of Defense Technology, Changsha, China\\
$^2$ College of Computer Science and Technology, National University of Defense Technology, Changsha, China\\
$^3$School of Artifcial Intelligence, Anhui University, Hefei, China\\
\emails
\{hyn241513, sihangjoe\}@gmail.com,\\
liangke200694@126.com,
mly\_edu@163.com,
xschenranker@gmail.com,
xuke@ahu.edu.cn,\\
\{wangsiwei13, xinwangliu, huang\_jian\}@nudt.edu.cn
}
\begin{document}

\maketitle

\begin{abstract}
Reasoning paths are reliable information in knowledge graph completion (KGC) in which algorithms can find strong clues of the actual relation between entities. However, in real-world applications, it is difficult to guarantee that computationally affordable paths exist toward all candidate entities. According to our observation, the prediction accuracy drops significantly when paths are absent. To make the proposed algorithm more stable against the missing path circumstances, we introduce soft reasoning paths. Concretely, a specific learnable latent path embedding is concatenated to each relation to help better model the characteristics of the corresponding paths. The combination of the relation and the corresponding learnable embedding is termed a soft path in our paper. By aligning the soft paths with the reasoning paths, a learnable embedding is guided to learn a generalized path representation of the corresponding relation. In addition, we introduce a hierarchical ranking strategy to make full use of information about the entity, relation, path, and soft path to help improve both the efficiency and accuracy of the model. Extensive experimental results illustrate that our algorithm outperforms the compared state-of-the-art algorithms by a notable margin. Our code will be released at https://github.com/7HHHHH/SRP-KGC.

\end{abstract}

\section{Introduction}

Knowledge graphs (KGs) have emerged as a foundational framework for organizing and utilizing structured information in mission-critical domains, including question answering~\cite{apply1,apply3}, recommendation systems~\cite{apply2}, and information retrieval~\cite{GraphRAG}. Structurally, KGs are composed of triples \((h, r, t)\), where \(h\) denotes the head entity, \(r\) specifies the semantic relationship and \(t\) identifies the tail entity. However, despite their practical importance, KGs often exhibit incompleteness. This inherent limitation underscores the importance of Knowledge Graph Completion (KGC) techniques, which play a pivotal role in automating the knowledge graph construction and validation processes.

Existing knowledge graph completion methods can be broadly categorized into two main categories: embedding-based methods~\cite{TransE,RotatE,TuckER} and text-based methods~\cite{SimKGC,GHN,CSProm-KG}. With the advent of language models, their advanced linguistic understanding capabilities have significantly improved the performance of text-based methods. Taking full advantage of the semantic relationships between candidate tail entities and the query, text-based approaches have gained widespread adoption due to their substantial improvements in accuracy.

Recent studies~\cite{Red,BertRL} have investigated the incorporation of reasoning path information into text-based knowledge graph completion, where reasoning paths serve as valuable indicators for predicting entity relationships, leading to significant improvements in prediction accuracy. However, our empirical analysis reveals that these algorithms exhibit a marked decrease in performance when reasoning paths are not available. Furthermore, through a detailed statistical evaluation, we found that approximately 82\% of the triples in the WN18RR test set and roughly 27\% of the triples in the FB15K-237 test set do not contain valid 2-hop or 3-hop reasoning paths. This observation suggests that a substantial portion of entities lack accessible reasoning paths that could be utilized for relation prediction. Consequently, this limitation severely constrains the performance ceiling of these methods. Moreover, prior path-based methods require reasoning path searches and ranking for all candidate tail entities to achieve high prediction accuracy, resulting in long testing times and limiting their practical application in real-world scenarios.

To address the aforementioned challenges, we propose a knowledge graph completion method based on soft reasoning paths (SRP-KGC). Specifically, the proposed soft reasoning paths are formed by combining relations and learnable embeddings. By assigning an independent learnable embedding to each type of relation and then aligning it with the paths of that relation, our approach enables the modeling of various path information corresponding to the same relation using soft reasoning paths. In cases where reasoning paths are missing, soft reasoning paths effectively fill the gaps, thereby enhancing the stability and robustness of the algorithm in such scenarios.

Additionally, to improve the scalability of the algorithm and mitigate the negative impact of extensive path searches on efficiency while maintaining the accuracy of the ranking, we propose a hierarchical ranking strategy. This approach utilizes a combination of relation, reasoning path, and soft reasoning path evaluation metrics to perform tiered filtering, effectively ensuring the scalability of the algorithm for test entities.
Our contributions are summarized as follows:
\begin{itemize}
    \item We identify an overlooked issue of performance degradation in path-based algorithms when paths are missing and propose a KGC method based on soft reasoning paths that enhances the algorithm's stability against the candidate entities whose path information is absent.
    \item We propose a hierarchical ranking method based on relations, reasoning paths, and soft reasoning paths, which alleviate the scalability defect of the path-based algorithm and enhances its practical value.
    \item Extensive experimental results demonstrate that the soft reasoning paths constructed based on trainable embeddings can effectively narrow the semantic gap between relations and their corresponding holistic reasoning paths, while enhancing the discriminative ability of relational representations in path discrimination.
\end{itemize}

\section{Related Work}
\subsection{Knowledge Graph Completion}
Existing methods for knowledge graph completion (KGC) fall into two categories: embedding-based and text-based.Embedding-based methods encode entities and relations as vectors. Translational models (e.g., TransE~\cite{TransE}, TransH~\cite{TransH}) are efficient but weak in modeling complex patterns. Tensor models like ComplEx~\cite{ComplexEF} handle diverse relations but scale poorly. Graph neural networks (e.g., CompGCN~\cite{CompGCN}) incorporate neighbor information to improve representations, though they require careful architecture design.Text-based methods leverage textual context. KG-BERT~\cite{KG-BERT} encodes triples as text for classification. SimKGC~\cite{SimKGC} and C-LMKE~\cite{C-LMKE} use contrastive learning for better discrimination. However, these methods often rely only on \((h, r)\) and candidate entity text similarity, ignoring richer auxiliary cues.
\subsection{Reasoning Path in KGC}
Reasoning is crucial for accurate knowledge graph completion (KGC). Unlike traditional embedding-based methods, reasoning path-based approaches capture higher-order relations by exploring paths that reflect semantic or logical connections. GraIL~\cite{GraIL} uses GNNs to assess path-relation relevance, BERTRL~\cite{BertRL} encodes reasoning paths and candidate triples with BERT, and ReDistLP~\cite{Red} aggregates multiple paths for prediction. These methods excel in inductive KGC but struggle when reasoning paths are missing or candidate triples are abundant.
\subsection{Prompt Tuning}
Prompt tuning~\cite{hou}, through the use of prompts, enables pre-trained language models (PLMs)~\cite{hou2} to achieve exceptional performance across various downstream tasks with minimal computational cost. CSProm-KG~\cite{CSProm-KG} is the first work to incorporate prompt tuning into KGC tasks. By applying prefix tuning in conjunction with GNNs, it effectively completes the KGC task under low-parameter conditions. AutoKG~\cite{autoKG} also explores the application of prompt engineering within the knowledge graph domain. A frequently overlooked aspect of prompt tuning is its capacity to learn general representations of data during the training process, a feature that our method leverages. This enables our model not only to handle specific tasks but also to extract and utilize general patterns from the data, thereby enhancing the model's generalization ability and overall performance.
\begin{figure*}[!htb]
    \centering
    \includegraphics[width=1\textwidth]{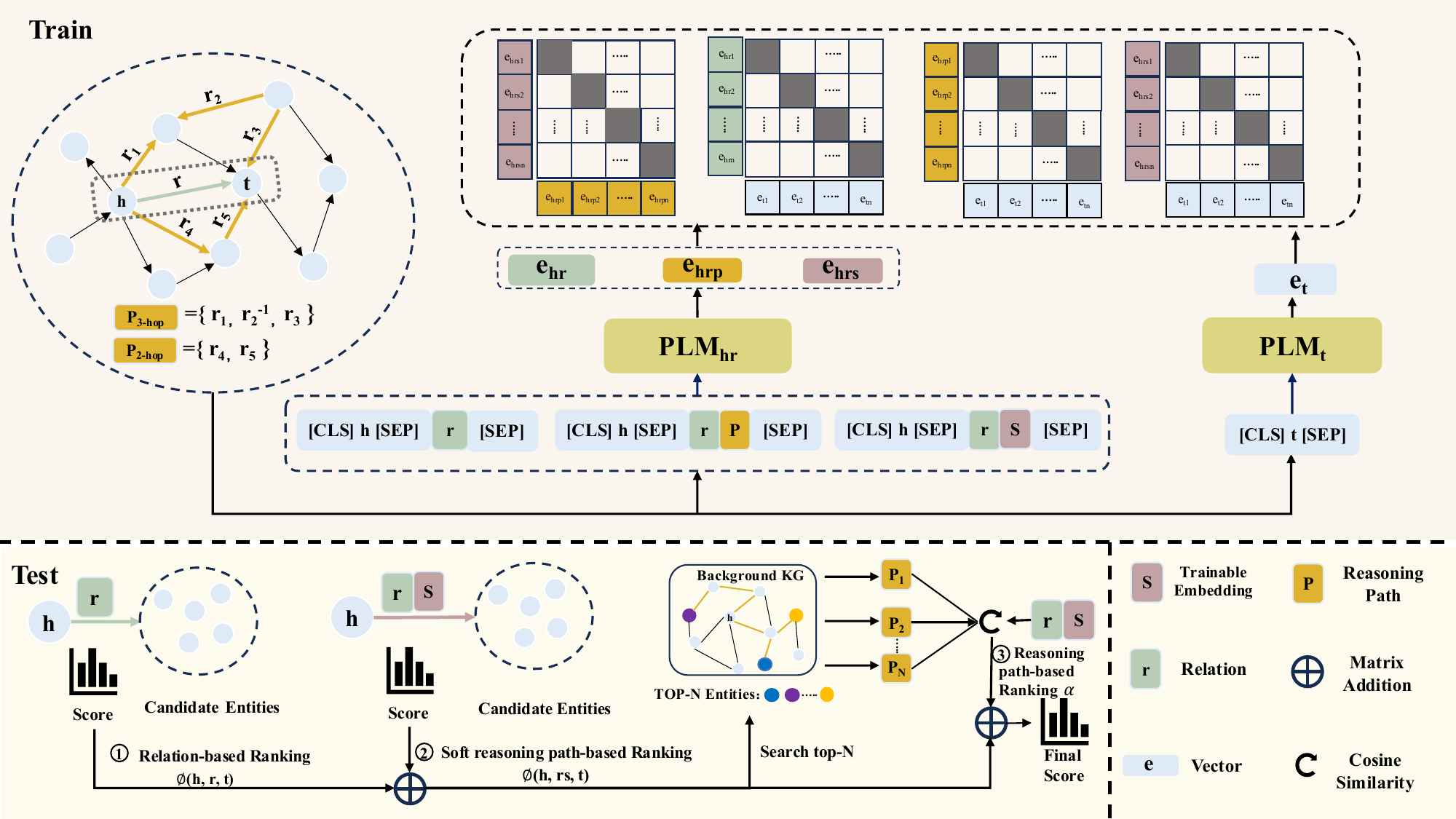}  
    \caption{SRP-KGC Framework: During the training process, we introduced three types of positive samples. By incorporating these diverse positive samples, the model's ability to understand reasoning paths was enhanced, while the soft reasoning path learns the generalized representation of reasoning paths. In the testing phase, we employed a hierarchical ranking strategy, combining information from entities, relations, soft reasoning paths, and reasoning paths to further improve the model's accuracy.}
    \label{fig:structure}
\end{figure*}
\section{Method}
\subsection{Problem Statement}\label{sec:3.1}
Given a knowledge graph \( G = \{(h, r, t) \mid h, t \in E, r \in R\} \), where \( E \) and \( R \) are the set of entities and relations of the KG, respectively. \( h \) and \( t \) are the head and tail entities, while \( r \) is the relation between them. The KGC task aims to predict the missing triples. In the entity ranking evaluation protocol, tail entity prediction \( (h, r, ?) \) ranks all entities based on \( h \) and \( r \), while head entity prediction \( (?, r, t) \) does the same. In this paper, we follow the SimKGC~\cite{SimKGC} setup and add an inverse triple \( (t, r^{-1}, h) \) for each triple \( (h, r, t) \), simplifying the task to only tail entity prediction.
\subsection{Network Framework Based on Contrastive Learning}\label{sec:3.2}
The proposed SRP-KGC method is based on a dual-encoder contrastive learning architecture and consists of three main components. First, we use multi-type positive samples for contrastive learning, introducing reasoning paths during the training phase to guide the model in enhancing its ability to discriminate reasoning paths. Next, we introduce soft reasoning paths, and by aligning soft reasoning paths with reasoning paths, we guide the model to learn generalized path representations of the corresponding relations to alleviate the issue of missing reasoning paths. Finally, during the testing phase, to fully utilize the information from entities, relations, reasoning paths, and soft reasoning paths, we introduce a hierarchical ranking strategy, combining multiple sources of information to further improve the accuracy of predictions.
\subsection{Multi-Type Positive Samples}\label{sec:3.2}
In the contrastive learning framework, we use three types of positive samples: relation positive samples, reasoning path positive samples, and soft reasoning path positive samples. Relation positive samples are triples (\( h, r, t \)) where the head and tail entities are directly related by relation \( r \). Reasoning path positive samples replace the direct relation with a reasoning path from \( h \) to \( t \), while soft reasoning path positive samples involve learning a generalized representation of the reasoning path through trainable embeddings, which will be explained in the next section.

Reasoning paths are the foundation of our approach. To ensure their generalization, we focus on relations and ignore entity information, represented as \( p = \{r_1, r_2, \dots\} \). Paths are classified based on the number of hops \( n \), with 2-hop and 3-hop paths being considered. We use a path constraint resource allocation algorithm from ~\cite{PCRA} to compute the confidence of each path and retain the highest-scoring ones. Additionally, we add the original relations as prefixes to each path to improve their expressiveness, creating a composite representation.
\begin{equation}
I_{r}^p =[CLS]h[SEP]r[SEP]
\end{equation}
\begin{equation}
I_{rp2}^p = [CLS]h[SEP]r p_2[SEP]
\end{equation}
\begin{equation}
I_{rp3}^p = [CLS]h[SEP]r p_3[SEP]
\end{equation}
In our framework, each relation or reasoning path, combined with the corresponding head entity, forms query texts (\( I_{r}^p \), \( I_{rp2}^p \), \( I_{rp3}^p \)). These query texts are paired with the correct tail entity \( t \) to generate positive samples. Then, these text pairs are processed through two BERT modules: the relation-aware module (\( Bert_{hr} \)) encodes the query text, generating embeddings \( e_{hr} \), \( e_{hrp2} \), and \( e_{hrp3} \). The entity-specific module (\( Bert_t \)) independently encodes the tail entity and generates the embedding \( e_t \).
\begin{equation}
\mathcal{L}_{\mathrm{hr\_t}}=\mathcal{L}(e_{hr},e_t),
\mathcal{L}_{\mathrm{hp\_t}}=\mathcal{L}(e_{hrp2},e_t)+\mathcal{L}(e_{hrp3},e_t)
\end{equation}
Here, \( \mathcal{L} \) represents the loss function, and its specific form will be introduced in detail later. 
\subsection{Soft Reasoning Paths}\label{sec:3.3}
To alleviate the issue of absent reasoning paths, we introduce soft reasoning paths. Specifically, for each relation \( r \) or inverse relation \( r^{-1} \), we append a trainable embedding \( \mathbf{S}_r \in \mathbb{R}^{d_{out} \times m} \) to it. By concatenating trainable embedding with the original relation representation, we construct soft reasoning paths that are capable of generalizing path semantics. During training, we design a contrastive learning objective to align the soft reasoning path embedding \( e_{hrs} \) with the encoded authentic reasoning paths \( e_{hrp2} \) and \( e_{hrp3} \) in vector space. This alignment guides the soft reasoning paths to learn generalized path patterns of relations from a limited set of path samples. This design allows soft reasoning paths to simulate latent reasoning logic through generalized representations during testing, even when reasoning paths are absent, thus significantly mitigating the impact of missing paths on prediction performance.

Specifically, we first construct sentence pairs \( I_{r}^p \). After token embedding, we append a trainable embedding \( \mathbf{S_r} \)  to the embedding vector of the relation. We define the soft reasoning path as \( I_{rs}^p \).
\begin{equation}
I_{rs}^p =[CLS]h[SEP]r S_r[SEP],
S_r  = W_{2}\cdot(\mathrm{ReLU}(W_{1}\cdot x_r))
\end{equation}
where \( x_r \in \mathbb{R}^{d_{in} \times m} \), \( m \) denotes the number of relations, and \( d_{in} \) represents the dimensionality of the trainable embeddings we define. \( W_1 \in \mathbb{R}^{d_h \times d_{in}} \) and \( W_2 \in \mathbb{R}^{d_{out} \times d_h} \) are trainable weight matrices. \( d_h \) denotes the dimensionality of the hidden layer, while \( d_{out} \) represents the dimensionality of the output layer. The output layer is typically expressed as \( l \times 768 \), where \( l \) is the number of trainable embeddings, and 768 is the default dimensionality of BERT input embeddings. A corresponding \( x_r \) is assigned to each relation. Specifically, if the knowledge graph contains 247 types of relations, there will be a total of \( 247 \times 2 \) instances of \( x_r \) (accounting for both forward and inverse relations). Notably, \( W_1 \) and \( W_2 \) are shared parameters across all relations. The soft reasoning path plays a role in learning the representations of reasoning paths to better capture complex reasoning information.
\begin{equation}
\mathcal{L}_{\mathrm{hrs\_t}}=\mathcal{L}(e_{hrs},e_t)
\end{equation}
\begin{equation}
\mathcal{L}_{\mathrm{hrs\_p}}=\mathcal{L}(e_{hrs},e_{hp2})+\mathcal{L}(e_{hrs},e_{hp3})
\end{equation}
Here, \( e_{hrs} \) is the result of encoding \( I_{rs}^p \) with \( {Bert_{hr}} \).
\subsection{Hierarchical Ranking}\label{sec:3.4}
During the testing phase, we predict the tail entities using the known head entities and relations. In this process, in addition to the relational information, we can also leverage the soft reasoning paths learned during training (i.e., \((h, r)\) and \((h, rs)\)). By employing a dual-encoder architecture, we preprocess all candidate tail entities, enabling efficient and rapid computation. Although reasoning paths are highly valuable, performing path searches for every candidate entity would incur substantial computational costs. For instance, in the Wikidata5M dataset, each triple contains 4,594,485 candidate tail entities, making exhaustive computation impractical. To address this challenge, our approach strikes a balance between computational cost and performance through a hierarchical ranking strategy. During the reasoning phase, we first perform a quick filtering using the relations and soft reasoning paths, and then conduct reasoning path searches only for the high-confidence candidate entities.
\begin{equation}
\text{Logits} = \phi(h, r, t) + \phi(h, rs, t), 
\hat{E} = \text{Top-N}( \text{Logits})
\end{equation}
Here, we define \( \phi(h, r, t) = \cos(\mathbf{e}_{hr}, \mathbf{e}_t) \in [-1, 1] \), and similarly, \( \phi(h, rs, t) = \cos(\mathbf{e}_{hrs}, \mathbf{e}_t) \in [-1, 1] \). Next, we select the top \( N \) candidate entities with the highest scores for the current triple \( (h, r) \), and perform path searches in the known graph between the head entity and these candidate entities. Here, \( N \) is a tunable ranking parameter that can be adjusted flexibly based on the characteristics of the dataset.
\begin{equation}
Path_2,Path_3 = \text{Search}(h, \hat{E})
\end{equation}
Here, we still limit the search to only 2-hop and 3-hop paths, i.e., \( Path_2 \) and \( Path_3 \). After combining the searched paths with the head entity and passing them through \( \text{Bert}_h \), we obtain the embeddings \( \mathbf{e}_{hp2} \) and \( \mathbf{e}_{hp3} \). Then, we calculate the similarity between these two vectors and \( \mathbf{e}_{hrs} \) by computing the cosine similarity, yielding a value \( \alpha \in [-1, 1] \). From these results, we select the one with the highest score.
\begin{equation}
\alpha = \max \left( \cos(\mathbf{e}_{hp2}, \mathbf{e}_{hrs}), \cos(\mathbf{e}_{hp3}, \mathbf{e}_{hrs}) \right)
\end{equation}
We add the obtained \( \alpha \) values to the high-confidence candidate entities in order to further optimize the results. This adjustment allows the model to prioritize the most relevant entities, improving the overall performance of the KGC task.
\begin{table*}[htbp]
  \centering
 \setlength{\tabcolsep}{5pt}
    \small
    \begin{tabular}{c|cccc|cccc|cccc}
    \toprule
    \multirow{2}[4]{*}{Methods} & \multicolumn{4}{c|}{WN18RR}   & \multicolumn{4}{c|}{FB15k-237} & \multicolumn{4}{c}{Wikidata5M-Trans} \\
\cmidrule{2-13}          & MRR   & Hits@1 & Hits@3 & Hits@10 & MRR   & Hits@1 & Hits@3 & Hits@10 & MRR   & Hits@1 & Hits@3 & Hits@10 \\
    \midrule
    \multicolumn{13}{c}{Embedding-based methods} \\
    \midrule
    TransE & 24.3  & 4.3   & 44.1  & 53.2  & 27.9  & 19.8  & 37.6  & 44.1  & 25.3  & 17.0  & 31.1  & 39.2  \\
    ComplEx & 44.9  & 40.9  & 46.9  & 53.0  & 27.8  & 19.4  & 29.7  & 45.0  & 28.2  & 22.6  & -     & 39.7  \\
    RotatE & 47.6  & 42.8  & 49.2  & 57.1  & 33.8  & 24.1  & 37.5  & 53.3  & 29.0  & 23.4  & 32.2  & 39.0  \\
    ConvE & 45.6  & 41.9  & 47.0  & 53.1  & 31.2  & 22.5  & 34.1  & 49.7  & -     & -     & -     & - \\
    CompGCN & 48.1  & 44.8  & 49.2  & 54.8  & 35.5  & 26.4  & 39.0  & 53.5  & -     & -     & -     & - \\
    TuckER & 47.0  & 44.3  & 48.2  & 52.6  & 35.8  & 26.6  & 39.4  & 54.4  & -     & -     & -     & - \\
    CompoundE & 49.2  & 45.2  & 51.0  & 57.0  & 35.0  & 26.2  & 39.0  & 54.7  & -     & -     & -     & - \\
    KPACL & 52.7  & 48.2  & 54.7  & 61.3  & 36.0  & 26.6  & 39.5  & \underline{54.8}  & -     & -     & -     & - \\
    RotatE-VLP & 49.8  & 45.5  & 51.4  & 58.2  & \underline{36.2}  & 27.1  & 39.7  & 54.2  & -     & -     & -     & - \\
    \midrule
    \multicolumn{13}{c}{Text-based methods} \\
    \midrule
    KG-BERT & 21.6  & 4.1   & 30.2  & 52.4  & -     & -     & -     & 42.0  & -     & -     & -     & - \\
    StAR  & 40.1  & 24.3  & 49.1  & 70.9  & 29.6  & 20.5  & 32.2  & 48.2  & -     & -     & -     & - \\
    KG-S2S & 57.4  & 53.1  & 59.5  & 66.1  & 33.6  & 25.7  & 37.3  & 49.8  & -     & -     & -     & - \\
    C-LMKE & 61.9  & 52.3  & 67.1  & 78.9  & 30.6  & 21.8  & 33.1  & 48.4  & -     & -     & -     & - \\
    SimKGC & 67.1  & 58.7  &\underline{73.1}  & 81.7  & 33.3  & 24.6  & 36.2  & 51.0  & 35.3  & 30.1  & 37.4  & 44.8  \\
    CSProm-KG & 57.5  & 52.2  & 59.6  & 67.8  & 35.8  & 26.9  & 39.3  & 53.8  & \underline{38.0}  & \underline{34.3}  & \underline{39.9}  & 44.6  \\
    LP-BERT & 48.2  & 34.3  & 56.3  & 75.2  & 31.0  & 22.3  & 33.6  & 49.0  & -     & -     & -     & - \\
    GS-KGC &  -  & 34.6  & 51.6  &  - & -   & \underline{28.0}  & \underline{42.6}  & -  & -     & -     & -     & - \\
    GHN   & \underline{67.8}  & \underline{59.6}  & 71.9  & \underline{82.1}  & 33.9  & 25.1  & 36.4  & 51.8  & 36.4  & 31.7  & 38.0  & \underline{45.3}  \\
    \midrule
    SRP-KGC   & \textbf{ 70.5 } & \textbf{ 63.6 } & \textbf{ 74.4 } & \textbf{ 83.1 } & \textbf{ 43.1 } & \textbf{ 35.3 } & \textbf{ 46.1 } & \textbf{ 58.5 } & \textbf{ 40.9 } & \textbf{ 36.6 } & \textbf{ 43.0 } & \textbf{ 48.8 } \\
    \bottomrule
    \end{tabular}%
      \caption{Main results on WN18RR,FB15k-237 and Wikidata5M-Trans  datasets. Bold numbers represent the best and underlined numbers represent the second best.}
  \label{tab:sota}%
\end{table*}%

\subsection{Loss Function}\label{sec:3.5}
In the training process, to further enhance the generalizability of the knowledge learned by the soft reasoning paths, inspired by~\cite{lossbase}, we improve the InfoNCE loss function. We extend InfoNCE~\cite{loss} to handle multiple positive samples simultaneously by maximizing the likelihood of these positive samples, thus integrating shared semantic information. This modification allows the model to better capture diverse patterns in the data, improving its performance on KGC tasks.
\begin{equation}
\mathcal{L}=-\frac1{|P|}\sum_{r*\in P}\log\frac{e^{\phi\left(h,r*,t\right)/\tau}}{\sum_{i=1}^{|\mathcal{N}|}e^{\phi\left(h,r*,t_i\right)/\tau}}
\end{equation}
Here, \( P \) is the set of all previously mentioned positive samples, that is, the relation \( r \), 2-hop path \( rp_2 \), 3-hop path \( rp_3 \), and the soft reasoning path \( rs \). At the same time, we retain the temperature parameter \( \tau \) to balance the importance between the samples. In addition to the in-batch negative samples, we do not introduce any additional negative samples. 
\begin{equation}
\mathcal{L}_{all}=w_{1}\mathcal{L}_{\mathrm{hr\_t}}+w_{2}\mathcal{L}_{\mathrm{hp\_t}}+w_{3}\mathcal{L}_{\mathrm{hrs\_t}}+w_{4}\mathcal{L}_{\mathrm{hrs\_p}}
\end{equation}
Where \( w_{i} \) is tunable hyper-parameters for adapting to speciﬁc knowledge graph characteristics.

\section{Experiments}
In this section, we evaluate the overall performance of SRP-KGC and the effectiveness of its individual modules. The experiments aim to answer the following four research questions:
\begin{itemize}
    \item RQ1. How does the proposed SRP-KGC perform compared to the state-of-the-art methods under both transductive and inductive settings? (see Section \ref{sec:4.2})
    \item RQ2. Will the introduction of soft paths improve the discriminability of the reasoning path embedding? (see Section \ref{sec:4.3})
    \item RQ3. How does the soft reasoning path perform when reasoning paths are missing or present? (see Section \ref{sec:4.4})
    \item RQ4. How does hierarchical ranking work? Is it effective? (see Section \ref{sec:4.6})
\end{itemize}  
\subsection{Experimental Settings}\label{sec:4.1}
We evaluated our method on three commonly used datasets: WN18RR, FB15k-237, and Wikidata5M-Trans. Detailed information about these datasets is shown in Table \ref{tab:dataset}. During the evaluation on these datasets, the candidate entities included all entities in the respective datasets. In addition, ~\cite{GraIL} extracted four inductive versions (v1,v2,v3,v4) of datasets for both WN18RR and FB15k-237. When testing on these inductive datasets, we followed the conventional setup and used only 50 candidate entities that included the target tail entity for fair comparison. Due to space constraints, the detailed descriptions of the inductive datasets are provided in the appendix. 
\begin{table}[htbp]
  \centering
  \footnotesize 
  \setlength{\tabcolsep}{2pt} 
    \begin{tabular}{cccccc}
    \hline
    Dataset & \# Ent & \# Rel & \# train & \# valid & \# test \\
    \hline
    WN18RR & 40,943 & 11    & 86,835 & 3,034 & 3,134 \\
    FB15k-237 & 14,541 & 237   & 272,115 & 17,535 & 20,466 \\
    Wikidata5M-Trans & 4,594,485 & 822   & 20,614,279 & 5,163 & 5,163 \\
    \hline
    \end{tabular}%
      \caption{Statistics of the datasets.}
  \label{tab:dataset}%
\end{table}%

We adopted the text-based model SimKGC~\cite{SimKGC} as our baseline, retaining the BERT parameter settings from the original paper. Our implementation was built using PyTorch. Hyperparameters \( w_i \) were optimized via grid search over the set \( \{ 0.2, 0.4, 0.6, 0.8, 1 \} \). All experiments ran on 4 NVIDIA RTX 4090 24GB GPUs. Evaluation used four automated metrics:MRR: Mean reciprocal rank of test triples; Hit@\( k \): Proportion of correct entities in top-\( k \) predictions (\( k = 1, 3, 10 \)). The detailed hyperparameters can be found in the appendix.

\subsection{Performance Comparison with SOTA Method}\label{sec:4.2}
In this study, we conducted a comparative analysis of SRP-KGC, comparing it with both embedding-based and text-based approaches. The embedding-based methods include TransE~\cite{TransE}, ComplEx~\cite{ComplexEF}, RotatE~\cite{RotatE}, ConvE~\cite{ConvE}, TuckER~\cite{TuckER}, CompoundE~\cite{CompoundE}, KRACL~\cite{KRACL}, and RotatE-VLP~\cite{RotatE-VLP}. On the other hand, the text-based methods  include KG-BERT~\cite{KG-BERT}, StAR~\cite{StAR}, KG-S2S~\cite{KG-S2S}, C-LMKE~\cite{C-LMKE}, SimKGC~\cite{SimKGC}, CSProm-KG~\cite{CSProm-KG}, LP-BERT~\cite{LP-BERT}, GS-KGC~\cite{GSKGC} and GHN~\cite{GHN}.

The main results are summarized in Table \ref{tab:sota}. Several conclusions can be drawn from these findings. Firstly, our method outperforms previous works across all metrics on the three datasets. Specifically, on the WN18RR dataset, our SRP-KGC improves the MRR and Hits@1 metrics by 4\% and 6.7\%, respectively. On Wikidata5M-Trans, it improves by 7.6\% and 6.7\%. Notably, on FB15k-237, our SRP-KGC improves by 19\% and 26.1\%. These results indicate that our SRP-KGC method demonstrates strong competitiveness in knowledge graphs with both sparse and dense topologies, as well as in large-scale knowledge graphs.
\begin{table}[htbp]
  \centering
  \footnotesize 
    \begin{tabular}{c|ccccc}
    \toprule
    \multirow{2}[4]{*}{Methods} & \multicolumn{5}{c}{WN18RR\_ind} \\
\cmidrule{2-6}          & V1    & V2    & V3    & V4    & AVG \\
    \midrule
    GraIL & 82.4  & 78.6  & 58.4  & 73.4  & 73.2  \\
    SimKGC & 95.8  & 97.2  & \textbf{96.2} & 97.4  & 96.7  \\
    GLAR  & 93.6  & 94.7  & 93.3  & 92.4  & 93.5  \\
    SRP-KGC   & \textbf{97.8} & \textbf{98.9} & 96.1  & \textbf{98.4} & \textbf{ 97.8 } \\
    \bottomrule
    \end{tabular}%
      \caption{The Hits@10 of WN18RR under inductive scenario. The optimal values of each metric are marked in blod.}
  \label{tab:ind_1}%
\end{table}%

To further explore the generalization capability of our method, we conducted experiments under the inductive KGC setting. The datasets used include WN18RR (v1, v2, v3, v4) and FB15k-237 (v1, v2, v3, v4), which were extracted by~\cite{GraIL}. Due to space constraints, we compared SRP-KGC with the following three approaches: GraIL~\cite{GraIL}, which  is one of the most classic methods for completing KGC tasks using reasoning paths; SimKGC~\cite{SimKGC}, which has a similar structure; and GLAR~\cite{GLAR}, the current state-of-the-art method. For a fair comparison, we adopted the experimental setup of GraIL, retaining only 50 candidate entities containing the target tail entity by default, and used Hits@10 as the evaluation metric. Tables \ref{tab:ind_1} and Tables \ref{tab:ind_2} present the experimental results on these two datasets. 
\begin{table}[htbp]
  \centering
  \footnotesize 
  \begin{tabular}{c|ccccc}
    \toprule
    \multirow{2}[4]{*}{Methods} & \multicolumn{5}{c}{FB15k-237\_ind} \\
    \cmidrule{2-6}          
    & V1    & V2    & V3    & V4    & AVG \\
    \midrule
    GraIL & 64.2  & 81.8  & 95.7  & 89.3  & 82.7  \\
    SimKGC & 90.5  & 93.2  & 91.0  & 90.1  & 91.2  \\
    GLAR  & 91.3  & 96.6  & 96.0  & \textbf{96.4} & 95.1  \\
    SRP-KGC & \textbf{96.3} & \textbf{98.1} & \textbf{96.2} & 95.3 & \textbf{96.5} \\
    \bottomrule
  \end{tabular}
  \caption{The Hits@10 of FB15k-237 under the inductive scenario. The optimal values of each metric are marked in bold.}
  \label{tab:ind_2}
\end{table}

The experiments demonstrate that SRP-KGC improves upon the best-performing method by 1.1\% and 1.5\%, respectively, in these two tasks. Through this approach, the model is able to effectively capture the underlying patterns of reasoning paths, demonstrating strong generalization ability even when handling previously unseen entities.
\subsection{Ability to Comprehend the Reasoning Path}\label{sec:4.3}
\begin{figure}[htbp]
    \centering
\includegraphics[width=0.4\textwidth]{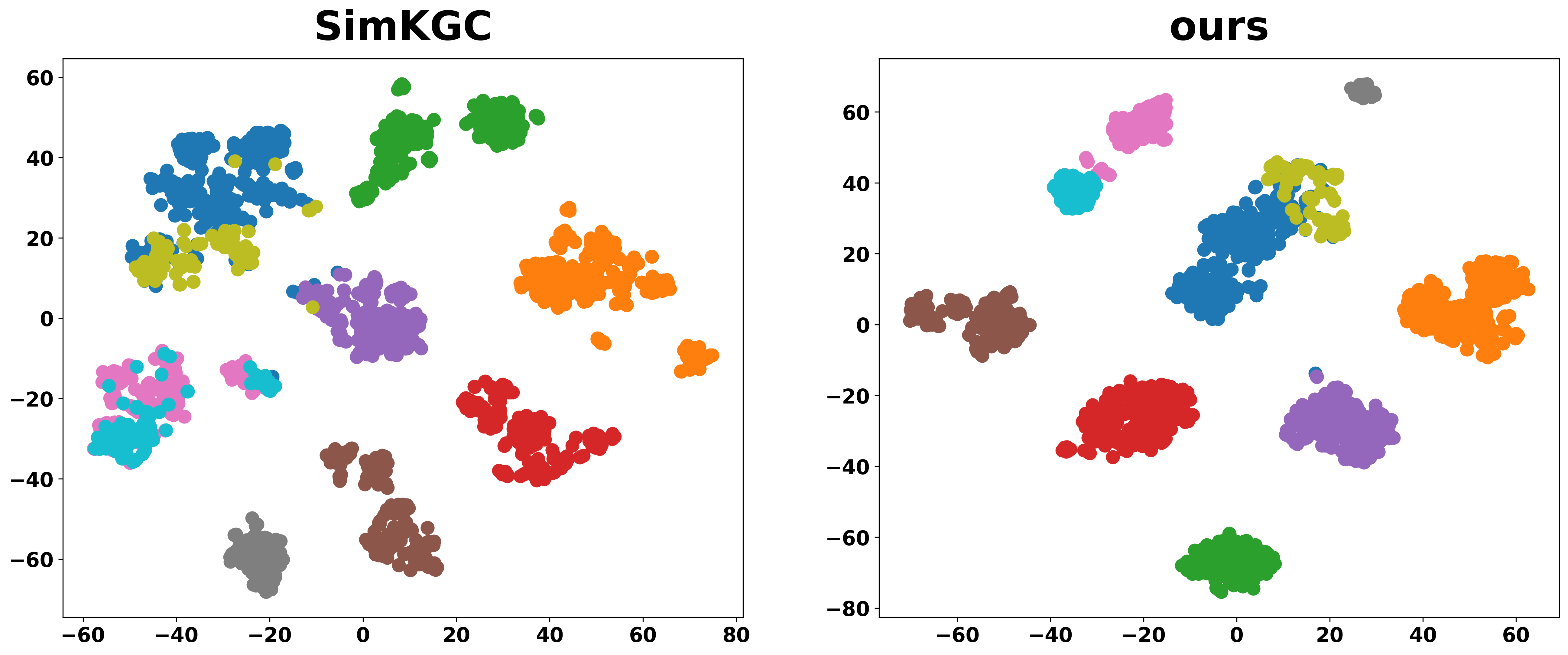}  
    \caption{Visualization of embeddings with the different head entities and relations using t-SNE under the settings of SimKGC and SRP-KGC. In the visualization, points with the same color represent embeddings that share the same target tail entity. }
    \label{fig:embedding}
\end{figure}
We compared the discriminative features obtained from different head entities pointing to the same tail entity through different paths and validated the effectiveness of incorporating multiple types of positive samples into the training process to enhance the model's ability to understand reasoning paths. In the experiment, we selected 10 tail entities that are highly relevant to triples from the FB15k-237 test set. For these tail entities and their related triples, we performed path searches and combined the resulting paths with their corresponding head entities. Subsequently, we encoded these combinations using BERT models trained with SimKGC and SRP-KGC. We visualized the encoded outcomes using t-SNE in Figure \ref{fig:embedding}. The visualization shows that after training with SRP-KGC, the embeddings for the same tail entity are significantly closer in the embedding space, demonstrating better feature discriminability.
\subsection{Effectiveness of Soft Reasoning Paths}\label{sec:4.4}
To validate the effectiveness of the soft reasoning path, we designed a series of comparative experiments, focusing on its performance in different scenarios. Specifically, we conducted comparisons under two conditions based on the existence of reasoning paths, and analyzed the impact of introducing the soft reasoning path:
\begin{table}[htbp]
  \centering
  \small

    \begin{tabular}{c|cccc}
    \toprule
    Testing settings & MRR   & Hits@1   & Hits@3   & Hits@10 \\
    \midrule
    R     & 26.9  & 18.0  & 29.6  & 44.1 \\
    RS    & \textbf{33.2} & \textbf{24.4} & \textbf{36.5} & \textbf{51.0} \\
    \bottomrule
    \end{tabular}%
  \caption{For the comparison of results in the absence of reasoning paths, R represents testing using relationships, while RS represents testing using soft reasoning paths.}
  \label{tab:w/o}%
\end{table}%

Without reasoning paths: In this scenario, previous models rely solely on the direct relationships within the triples for prediction, which fails to provide more effective information. Our approach introduces soft reasoning paths, and we conduct a comparative analysis. We collected 5,560 triplets from the FB15k-237 test set that do not have reasoning paths, and performed a separate analysis. As shown in Table \ref{tab:w/o}, the proposed Soft reasoning paths, in the case of missing paths, showed improvements over traditional methods by 23.4\%, 35.6\%, 23.3\%, and 15.6\% on the MRR, Hits@1, Hits@3, and Hits@10 metrics, respectively. These results demonstrate that Soft reasoning paths, by learning the representation of the same relationship under different paths, effectively alleviate the issue of missing reasoning paths in KGC tasks.

With reasoning paths: In this scenario, existing path-based methods determine the target tail entity by the correlation between the reasoning path and the target relation. However, reasoning paths are often stacks of relationships, resulting in a significant semantic gap from the target relation. To alleviate this, we compared the correlation between reasoning paths and relations, as well as the correlation between reasoning paths and soft reasoning paths, to evaluate the role of soft reasoning paths in reducing the semantic gap.

Replacing the soft reasoning path with relationships and applying it to the final step of our proposed hierarchical ranking strategy is an effective comparative method. We conducted comparisons on the structurally relatively dense FB15k-237 dataset, as shown in Table \ref{tab:rerank-com}, indicate that using Soft reasoning paths, compared to using relationships, led to improvements of 23.3\%, 30.5\%, 23.2\%, and 15.9\% in MRR, Hit@1, Hit@3, and Hit@10, respectively.
\begin{table}[htbp]
  \centering
  \small
    \begin{tabular}{c|cccc}
    \toprule
    Rank settings & MRR   & Hits@1 & Hits@3 & Hits@10 \\
    \midrule
    R     & 33.8  & 26.2  & 36.1  & 48.5 \\
    RS    & \textbf{41.7} & \textbf{34.2} & \textbf{44.5} & \textbf{56.2} \\
    \bottomrule
    \end{tabular}%
      \caption{The comparison of results during the ranking phase using relationships and soft reasoning paths is as follows: R represents relationships, and RS represents Soft reasoning paths.}
  \label{tab:rerank-com}%
\end{table}%

To further validate the effectiveness of soft reasoning paths, we collected 14,806 triples with reasoning paths and searched for their 2-hop and 3-hop paths. We performed relation prediction using both relations and soft reasoning paths for these paths. 
Specifically, the embedding vectors \( \mathbf{e}_r \)  of the relations involved in these triples, as well as the soft reasoning paths \( \mathbf{e}_{rs} \) corresponding to each relation, were computed. Subsequently, we encoded the embeddings of these reasoning paths using the same encoder to obtain  \( \mathbf{e}_p \). Finally, we calculated the similarities between \( \mathbf{e}_p \) and \( \mathbf{e}_r \), as well as between \( \mathbf{e}_p \)  and \( \mathbf{e}_{rs} \) and evaluated the results within their respective sets. As shown in Table \ref{tab:w path}, compared to using only relations, the use of soft reasoning paths improved the Hits@10, F1, and ROC-AUC metrics by 4.6\%, 4.8\%, and 2.1\%, respectively. Furthermore, if soft reasoning paths were not used during training, the corresponding improvements in the metrics would be 15.6\%, 94.7\%, and 6.3\%, respectively.
\begin{table}[htbp]
  \centering
   \fontsize{8}{10}\selectfont
    \begin{tabular}{c|c|ccc}
    \toprule
    Training Settings & Testing Settings & Hits@10  & F1    & ROC-AUC \\
    \midrule
    w/o RS & R & 74.5  & 19.0    & 49.5 \\
    w RS & R & 82.3  & 35.3  & 51.5 \\
    w RS & RS  & \textbf{86.1} & \textbf{37.0} & \textbf{52.6} \\
    \bottomrule
    \end{tabular}%
      \caption{Relation prediction performance across training and testing configurations. R represents relationships, and RS represents soft reasoning paths.}
  \label{tab:w path}%
\end{table}%

\subsection{Case Study}\label{sec:4.6}

\begin{table}[htbp]
  \centering
  \small  
  \setlength{\tabcolsep}{4pt}
    \begin{tabular}{@{}cccc@{}}  
    \toprule
    \multicolumn{3}{c}{Contact, language film, English Language} & Answer \\
    \cmidrule{1-3}    
    information & Top 3 candidate entities & probabilities & Rank \\
    \midrule
    \multirow{3}[2]{*}{(h,r)} & Greek Language & 0.547  & \multirow{3}[2]{*}{7} \\
          & Japanese Language & 0.543  &  \\
          & Hebrew Language & 0.530  &  \\
    \midrule
    \multirow{3}[2]{*}{(h,r)+(h,rs)} & Japanese Language & 1.034  & \multirow{3}[2]{*}{2} \\
          & \textbf{English Language} & 1.025  &  \\
          & Greek Language & 1.021  &  \\
    \midrule
    \multirow{3}[2]{*}{(h,r)+(h,rs)+(p)} & \textbf{English Language} & 1.458  & \multirow{3}[2]{*}{1} \\
          & Japanese Language & 1.262  &  \\
          & Greek Language & 1.249  &  \\
    \bottomrule
    \end{tabular}%
      \caption{The rankings and scores predicted by the model under different information conditions. The target entity is indicated in bold.}
  \label{tab:addlabel}%
\end{table}%

To further illustrate the effectiveness of the hierarchical ranking, we selected ``Contac'' as the head entity and ``language film'' as the relation for a prediction experiment. Specifically: Using \( (h, r) \) (head entity and relation) as the query, we calculated the similarity with all candidate entities, resulting in a rank of 7. Next, we added \( (h, rs) \) (head entity and soft reasoning path) as the query and performed similarity calculations again, improving the rank to 2. Finally, we conducted a search for the reasoning path and calculated the similarity between the reasoning paths and soft reasoning paths. Adding this score to the original score further improved the final rank to 1. This process demonstrates that integrating multiple types of information can effectively improve the accuracy of the model's predictions.
\section{Limitations}
Although SRP-KGC enhances KGC tasks by introducing soft reasoning paths, this leads to increased computational demands during ranking. To assess this, we examined the trade-off between performance gains and computational costs.Comparing SRP-KGC with BERTRL and SimKGC, SRP-KGC strikes the best balance between speed and accuracy. BERTRL takes 60 seconds per batch for a 3.5 MRR improvement, while SRP-KGC achieves an 8.1 MRR boost in just 15 seconds. Despite SRP-KGC requiring five times more processing time than SimKGC (which has a 0.8 MRR improvement), it offers ten times greater performance gains. The results show that SRP-KGC effectively enhances model precision through strategic computational allocation, surpassing BERTRL in efficiency and SimKGC in effectiveness. (The BERTRL results are from our reproduced experiments. We used our hierarchical ranking strategy to handle many candidate entities; without it, testing would take about 36 minutes per batch.)
\begin{table}[htbp]
  \centering
  \small  
    \begin{tabular}{@{}ccc@{}}  
    \toprule
    \multicolumn{3}{c}{Ranking Time Per Batch (512)} \\
    \midrule
    Methods & Time  & Ability (MRR) \\
    \midrule
    SimKGC & ~3s   & 32.8$\to$33.6 \\
    BERTRL & ~60s  & 32.0$\to$35.5 \\
    SRP-KGC & ~15s  & 33.6$\to$41.7 \\
    \bottomrule
    \end{tabular}%
      \caption{Comparison of ranking time and performance with SimKGC and BERTRL in FB15k-237.}
  \label{tab:time}%
\end{table}%

\section{Conclusion}
This paper proposes the SRP-KGC, which effectively alleviate issues such as missing reasoning paths, semantic gaps, and scalability in existing KGC tasks. By introducing learnable embeddings to construct soft reasoning paths and employing a hierarchical ranking strategy to fully leverage the available information, SRP-KGC significantly outperforms existing methods across multiple datasets, demonstrating its potential in large-scale KGC tasks. Although there is an increase in computational overhead, the substantial performance improvement indicates clear advantages of the method. Future research will focus on optimizing computational efficiency and further reducing time costs to enhance the practical applicability of the method.


\section*{Acknowledgments}
This work was supported by the Huxiang Young Talents in Science and Technology Innovation Project (No. 2024RC3148).

\bibliographystyle{named}
\bibliography{ijcai25}
\newpage

\appendix
\section{Appendix}
\subsection{Hyperparameters}\label{sec:a.1}
The Table \ref{tab:hyperparameter} presents the hyperparameters for our proposed SRP-KGC model across three datasets: WN18RR, FB15k-237, and Wikidata5M-Trans. It lists various hyperparameters, including the input dimension \(d_{in}\), hidden layer dimension \(d_{h}\), output dimension \(d_{out}\), trainable embedding number \(l\), relation number \(m\),learning rate, learning rate scheduler, and warmup steps. Additionally, other critical training parameters are provided, such as the initial temperature, number of epochs, batch size, gradient clipping value, and maximum tokens. Notably, the learning rates vary across datasets, and the weight vector \(w_{i}\) differs for the FB15k-237 dataset. These hyperparameter choices are designed to optimize model performance across different tasks and datasets.
\begin{table}[htbp]
  \centering
  \scriptsize  
  \caption{Hyperparameters for our proposed SRP-KGC  model.}
    \begin{tabular}{cccc}
    \toprule
    Hyperparameters & WN18RR & FB15k-237 & Wikidata5M-Trans \\
    \midrule
    \(d_{in}\)   & 144   & 144   & 144 \\
    \(d_{h}\)     & 72    & 72    & 72 \\
    \(d_{out}\)  & $l \times 768$   & $l \times 768$   & $l \times 768$ \\
    $l$           &10 &8 &8\\
    $m$ &22 & 474 & 1644\\
    Learning rate & 5e-5  & 1e-5  & 3e-5 \\
    LR Scheduler & Linear Warmup & Linear Warmup & Linear Warmup \\
    Warmup steps & 400   & 400   & 400 \\
    Initial temperature & 0.05  & 0.05  & 0.05 \\
    Epochs & 100   & 10    & 1 \\
    Batch size & 512   & 512   & 512 \\
    Gradient clipping & 10    & 10    & 10 \\
    Max tokens & 50    & 50    & 50 \\
    \(w_{i}\) & [1, 1, 1, 1] & [1, 1, 1, 0.2] & [1, 1, 1, 1] \\
    \bottomrule
    \end{tabular}%
  \label{tab:hyperparameter}%
\end{table}%
\subsection{Inductive datasets}\label{sec:a.2}
Due to space limitations in the main text, we present the relevant information for the datasets used in the inductive setting in Table \ref{tab:ind}. Each inductive dataset has four versions, with progressively increasing sizes.
\begin{table}[htbp]
  \centering
  \small
  \caption{Statistics of WN18RR-ind and FB15k237-ind datasets. \#R, \#E and \#T are the numbers of relations, entities and triples.}
    \begin{tabular}{cccccccc}
    \toprule
          &       & \multicolumn{3}{c}{WN18RR-ind} & \multicolumn{3}{c}{FB15k237-ind} \\
          &       & \#R   & \#E   & \#T   & \#R   & \#E   & \#T \\
    \midrule
    \multirow{2}[2]{*}{v1} & train & 9     & 2746  & 6678  & 183   & 2000  & 5226 \\
          & test  & 9     & 922   & 1991  & 146   & 1500  & 2404 \\
    \midrule
    \multirow{2}[2]{*}{v2} & train & 10    & 6954  & 18968 & 203   & 3000  & 12085 \\
          & test  & 10    & 2923  & 4863  & 176   & 2000  & 5092 \\
    \midrule
    \multirow{2}[2]{*}{v3} & train & 11    & 12078 & 32150 & 218   & 4000  & 22394 \\
          & test  & 11    & 5084  & 7470  & 187   & 3000  & 9137 \\
    \midrule
    \multirow{2}[2]{*}{v4} & train & 9     & 3861  & 9842  & 222   & 5000  & 33916 \\
          & test  & 9     & 7208  & 15157 & 204   & 3500  & 14554 \\
    \bottomrule
    \end{tabular}%
  \label{tab:ind}%
\end{table}%
\subsection{The effectiveness of hierarchical ranking}\label{sec:a.3}
The comparison of integrating different types of information for knowledge graph completion clearly shows the effectiveness of hierarchical ranking. As demonstrated by the results, the baseline model, which uses only the head entity and relation \( (h, r) \), shows solid performance on both the WN18RR and FB15K-237 datasets. However, when additional contextual information is introduced, such as the soft reasoning path  \( (h, rs) \), the performance improves, with increases observed in both MRR and Hits@1 metrics.

Notably, incorporating the reasoning path \( (p) \) alongside the previous two types of information provides the most significant improvement. This is especially evident on the FB15K-237 dataset, where the addition of reasoning paths substantially enhances the model's ability to rank the correct entities, evidenced by a large increase in both MRR and Hits@1. This suggests that the hierarchical integration of multiple layers of reasoning paths allows the model to better understand complex relationships within the knowledge graph, leading to more accurate and effective completion results.

The results underline the importance of hierarchical ranking in improving the model's performance, particularly in more challenging datasets like FB15K-237. The effectiveness of this approach highlights how layering different types of information can significantly enhance the model's ability to perform knowledge graph completion tasks, offering a compelling argument for the benefits of hierarchical ranking in such scenarios.
\begin{table}[htbp]
  \centering
  \caption{Comparison of the effects of integrating different types of information.}
    \begin{tabular}{c|cc|cc}
    \toprule
    \multirow{2}[4]{*}{Infomation} & \multicolumn{2}{c|}{WN18RR} & \multicolumn{2}{c}{FB15K-237} \\
\cmidrule{2-5}          & MRR   & Hits@1 & MRR   & Hits@1 \\
    \midrule
    (h,r) & 68.3  & 61.2  & 30.1  & 20.5 \\
    (h,r)+(h,rs) & 69.7  & 62.6  & 33.6  & 24.4 \\
    (h,r)+(h,rs)+(p) & \textbf{70.3}  & \textbf{63.4}  & \textbf{41.7}  & \textbf{34.2} \\
    \bottomrule
    \end{tabular}%
  \label{tab:addlabel}%
\end{table}%

\subsection{Ablation Study}\label{sec:a.4}
In this section, we focus on conducting ablation study for two important parameters: the number of trainable embeddings and the size of $N$ in hierarchical ranking.

\begin{table}[htbp]
  \centering
  \caption{The impact of trainable embeddings number on the FB15k-237 and WN18RR datasets}
    \begin{tabular}{c|cc|cc}
    \toprule
    embedding  & \multicolumn{2}{c|}{FB15K237} & \multicolumn{2}{c}{WN18RR} \\
    num   & MRR   & Hits@1   & MRR   & Hits@1 \\
    \midrule
    2     & 39.8  & 32.1  & 69.8  & 62.1 \\
    4     & 40.9  & 33.4  & 70.2  & 63.0 \\
    6     & 40.0  & 32.3  & 70.3  & 62.9 \\
    8     & \textbf{41.7} & \textbf{34.2} & 70.3  & 63.4 \\
    10    & 41.6  & 34.1  & \textbf{70.7} & \textbf{63.6} \\
    12    & 41.5  & 34.1  & 70.0  & 62.9 \\
    \bottomrule
    \end{tabular}%
  \label{tab:embedding_num}%
\end{table}%
We conducted ablation experiments with different numbers of trainable embeddings (i.e., {2, 4, 6, 8, 10, 12}) on the FB15k-237 and WN18RR datasets, evaluating the performance using MRR and Hits@1 (with a default rank number \( N \) of 100). As shown in Table \ref{tab:embedding_num}, we observed the following trends:FB15K237 Dataset:  
The performance improved with the increase in the number of embeddings from 2 to 8. Specifically, the MRR increased from 39.8 at 2 embeddings to 41.7 at 8 embeddings, and Hits@1 increased from 32.1 to 34.2. This indicates that adding more learnable embeddings enhanced the model's ability to capture and generalize path representations. However, the performance slightly decreased when the number of embeddings was increased beyond 8 (at 10 and 12 embeddings), where the MRR and Hits@1 values plateaued or showed slight declines. This suggests a diminishing return or potential overfitting as the embeddings become more complex and difficult to optimize effectively. WN18RR Dataset:
Similar patterns were observed, with the MRR and Hits@1 improving steadily from 2 embeddings (69.8 and 62.1, respectively) to 8 embeddings (70.3 and 63.4, respectively). Beyond 8 embeddings, the performance continued to improve slightly at 10 embeddings (MRR 70.7, Hits@1 63.6) but dropped slightly at 12 embeddings (MRR 70.0, Hits@1 62.9). Again, this trend indicates that while adding embeddings initially improves the model's performance, the gains become less significant and even start to reverse as the embedding space becomes too large to train effectively.

In summary, the experiment suggests that adding learnable embeddings enhances the model’s performance up to a point (8 embeddings for both datasets), beyond which further increases lead to diminishing returns or instability in training. This behavior may be due to the model becoming too complex or overfitting as the number of embeddings grows.
\begin{table}[htbp]
  \centering
  \caption{Impact of Top-N Filter Size on WN18RR and FB15k-237 Datasets.}
    \begin{tabular}{c|cc|cc}
    \toprule
    \multirow{2}[4]{*}{ N } & \multicolumn{2}{c|}{WN18RR} & \multicolumn{2}{c}{FB15K237} \\
\cmidrule{2-5}          & MRR   & Hits@1 & MRR   & Hits@1 \\
    \midrule
    0     & 69.7  & 62.6  & 33.6  & 24.4 \\
    100   & 70.3  & 63.4  & 41.7  & 34.2 \\
    200   & 70.4  & 63.5  & 42.5  & 34.8 \\
    300   & 70.4  & 63.5  & 42.9  & 35.1 \\
    400   & \textbf{70.5} & \textbf{63.6} & \textbf{43.1} & \textbf{35.3} \\
    \bottomrule
    \end{tabular}%
  \label{tab:rerank}%
\end{table}%

We conducted an ablation study to analyze the sensitivity of model performance to the Top-N filter size ($N$) across two benchmark datasets. As shown in Table \ref{tab:rerank}, we evaluated five configurations of $N \in \{0, 100, 200, 300, 400\}$, where $N=0$ serves as the baseline without entity filtering. Key observations include: WN18RR Dataset: Incremental gains in both MRR and Hits@1 were observed as $N$ increased from 0 to 400. The marginal improvements suggest that smaller candidate subsets (e.g., $N \leq 400$) sufficiently preserve high-quality entities in sparse knowledge graphs. FB15k-237 Dataset: Performance exhibited stronger dependency on $N$, with MRR improving by 28.3\%  and Hits@1 by 44.7\%  as $N$ increased from 0 to 400. Notably, 60\% of these gains were achieved at $N=100$, indicating that even moderate filtering significantly benefits dense knowledge graphs. This phenomenon aligns with our hypothesis that structural density amplifies the importance of selective entity retention during reasoning. In addition, we calculated that for every 100 increase in $N$, the ranking time increases by approximately 15 seconds per batch.

\subsection{Train time analysis}\label{sec:a.1}
In this section, we compared the time required to train one epoch between our method and SimKGC. Although our method incurs an increase in training time compared to other approaches, the performance improvement is substantial enough to justify the additional time cost. On the WN18RR and FB15K237 datasets, the training time per epoch for our method is 4 minutes and 27 seconds, and 14 minutes and 38 seconds, respectively, which is an increase of approximately 1 to 4 minutes compared to the SimKGC method. However, as training progresses, our method achieves scores of 70.5 and 43.1 on the ability evaluation metric (MRR), which represent improvements of 3.4 and 9.8 points, respectively, compared to SimKGC. Therefore, despite the higher time cost, the performance gains are significant and acceptable.
\begin{table}[htbp]
  \centering
  \scriptsize  
  \caption{Training time of one epoch and final performance compared to SimKGC}
    \begin{tabular}{@{}ccccc@{}}  
    \toprule
    \multicolumn{5}{c}{Training Time Per Epoch} \\
    \midrule
    Methods & WN18RR & Ability (MRR) & FB15K237 & Ability (MRR) \\
    \midrule
    SimKGC & 2m45s & 67.1  & 10m24s & 33.3 \\
    ours   & 4m27s & 70.5  & 14m38s & 43.1 \\
    \bottomrule
    \end{tabular}%
  \label{tab:addlabel}%
\end{table}%
\subsection{Trainable parameters}\label{sec:a.1}
\begin{table}[htbp]
  \centering
  \small
  \caption{Compare the trainable parameters with SimKGC}
    \begin{tabular}{cccc}
    \toprule
    \multicolumn{4}{c}{Trainable parameters} \\
    \midrule
    \multicolumn{1}{l}{Methods} & WN18RR & \multicolumn{1}{l}{FB15K-237} & \multicolumn{1}{l}{Wikidata5M-Trans} \\
    \midrule
    SimKGC & 218.0M & 218.0M & 218.0M \\
    ours  & 219.7M & 220.0M & 220.6M \\
    \bottomrule
    \end{tabular}%
  \label{tab:tp}%
\end{table}%

Our method introduces only a minimal increase in the number of trainable parameters. As shown in Table \ref{tab:tp}, we compared the number of parameters when the number of learnable embeddings is 8. On the WN18RR, FB15K-237, and Wikidata5M-Trans datasets, the number of trainable parameters for SimKGC and our method are 218.0M and 219.7M, 220.0M and 220.6M, respectively, with a very limited increase in the number of parameters. Therefore, our method provides significant performance improvements while maintaining minimal additional parameter overhead, demonstrating its high efficiency and optimization potential.
\subsection{Case Study}\label{sec:a.7}
In Table \ref{tab:cs1}-\ref{tab:cs6}, we present more examples of predictions made by SRP-KGC to help better understand the testing process of our model.
\begin{table}[htbp]
  \centering
    \fontsize{8}{10}\selectfont
  \caption{The rankings and scores predicted by the model for forward tail entity inference under varying information conditions are presented. The target entity is highlighted in bold.}
    \begin{tabular}{cccc}
    \toprule
    \multicolumn{3}{c}{Sandra Bernhard, profession person people, Actor-GB} & Answer \\
\cmidrule{1-3}    test set & Top3 candidate entites & probabilities & Rank \\
    \midrule
    \multirow{3}[2]{*}{(h,r)} & Spokesperson-GB & 0.548  & \multirow{3}[2]{*}{7} \\
          & Activism & 0.543  &  \\
          & Presenter-GB & 0.539  &  \\
    \midrule
    \multirow{3}[2]{*}{(h,r)+(h,rs)} & Activism & 1.051  & \multirow{3}[2]{*}{3} \\
          & Spokesperson-GB & 1.044  &  \\
          & \textbf{Actor-GB} & 1.004  &  \\
    \midrule
    \multirow{3}[2]{*}{(h,r)+(h,rs)+(p)} & \textbf{Actor-GB} & 1.526  & \multirow{3}[2]{*}{1} \\
          & Activism & 1.309  &  \\
          & Spokesperson-GB & 1.302  &  \\
    \bottomrule
    \end{tabular}%
  \label{tab:cs1}%
\end{table}%
\begin{table}[htbp]
  \centering
    \fontsize{8}{10}\selectfont
  \caption{The rankings and scores predicted by the model for forward tail entity inference under varying information conditions are presented. The target entity is highlighted in bold.}
    \begin{tabular}{cccc}
    \toprule
    \multicolumn{3}{c}{The Painted Veil, language film, French Language} & Answer \\
\cmidrule{1-3}    test set & Top3 candidate entites & probabilities & Rank \\
    \midrule
    \multirow{3}[2]{*}{(h,r)} & Persian Language & 0.578  & \multirow{3}[2]{*}{32} \\
          & Arabic Language & 0.573  &  \\
          & Hebrew Language & 0.559  &  \\
    \midrule
    \multirow{3}[2]{*}{(h,r)+(h,rs)} & Persian Language & 1.107  & \multirow{3}[2]{*}{21} \\
          & Arabic Language & 1.102  &  \\
          & Hebrew Language & 1.062  &  \\
    \midrule
    \multirow{3}[2]{*}{(h,r)+(h,rs)+(p)} & \textbf{French Language} & 1.553  & \multirow{3}[2]{*}{1} \\
          & Persian Language & 1.517  &  \\
          & Arabic Language & 1.512  &  \\
    \bottomrule
    \end{tabular}%
  \label{tab:cs2}%
\end{table}%
\begin{table}[htbp]
  \centering
    \fontsize{8}{10}\selectfont
  \caption{The rankings and scores predicted by the model for forward tail entity inference under varying information conditions are presented. The target entity is highlighted in bold.}
    \begin{tabular}{cccc}
    \toprule
    \multicolumn{3}{c}{Curly Howard, profession person people, Actor-GB} & Answer \\
\cmidrule{1-3}    test set & Top3 candidate entites & probabilities & Rank \\
    \midrule
    \multirow{3}[2]{*}{(h,r)} & Clown & 0.522  & \multirow{3}[2]{*}{5} \\
          & Screenwriter & 0.521  &  \\
          & Film Producer-GB & 0.519  &  \\
    \midrule
    \multirow{3}[2]{*}{(h,r)+(h,rs)} & Clown & 1.047  & \multirow{3}[2]{*}{2} \\
          & \textbf{Actor-GB} & 1.009  &  \\
          & Screenwriter & 0.995  &  \\
    \midrule
    \multirow{3}[2]{*}{(h,r)+(h,rs)+(p)} & \textbf{Actor-GB} & 1.487  & \multirow{3}[2]{*}{1} \\
          & Clown & 1.426  &  \\
          & Screenwriter & 1.412  &  \\
    \bottomrule
    \end{tabular}%
  \label{tab:cs3}%
\end{table}%
\begin{table}[htbp]
  \centering
    \fontsize{8}{10}\selectfont
  \caption{The rankings and scores predicted by the model for backward tail entity inference under varying information conditions are presented. The target entity is highlighted in bold.}
    \begin{tabular}{cccc}
    \toprule
    \multicolumn{3}{c}{Nas, artists genre music$^{-1}$, Hip hop music} & Answer \\
\cmidrule{1-3}    test set & Top3 candidate entites & probabilities & Rank \\
    \midrule
    \multirow{3}[2]{*}{(h,r)} & Jazz rap & 0.686  & \multirow{3}[2]{*}{9} \\
          & G-funk & 0.679  &  \\
          & Underground hip hop & 0.670  &  \\
    \midrule
    \multirow{3}[2]{*}{(h,r)+(h,rs)} & Jazz rap & 1.326  & \multirow{3}[2]{*}{7} \\
          & G-funk & 1.322  &  \\
          & Underground hip hop & 1.293  &  \\
    \midrule
    \multirow{3}[2]{*}{(h,r)+(h,rs)+(p)} & \textbf{Hip hop music} & 2.173  & \multirow{3}[2]{*}{1} \\
          & Jazz rap & 1.893  &  \\
          & G-funk & 1.889  &  \\
    \bottomrule
    \end{tabular}%
  \label{tab:cs4}%
\end{table}%
\begin{table}[htbp]
  \centering
    \fontsize{8}{10}\selectfont
  \caption{The rankings and scores predicted by the model for backward tail entity inference under varying information conditions are presented. The target entity is highlighted in bold.}
    \begin{tabular}{cccc}
    \toprule
    \multicolumn{3}{c}{Dick Clark, cause of death people$^{-1}$, Myocardial infarction} & Answer \\
\cmidrule{1-3}    test set & Top3 candidate entites & probabilities & Rank \\
    \midrule
    \multirow{3}[2]{*}{(h,r)} & Renal failure & 0.667  & \multirow{3}[2]{*}{20} \\
          & Cancer & 0.658  &  \\
          & Lung cancer & 0.655  &  \\
    \midrule
    \multirow{3}[2]{*}{(h,r)+(h,rs)} & Renal failure & 1.306  & \multirow{3}[2]{*}{17} \\
          & Cancer & 1.286  &  \\
          & Lung cancer & 1.269  &  \\
    \midrule
    \multirow{3}[2]{*}{(h,r)+(h,rs)+(p)} & \textbf{Myocardial infarction} & 1.707  & \multirow{3}[2]{*}{1} \\
          & Renal failure & 1.575  &  \\
          & Cancer & 1.555  &  \\
    \bottomrule
    \end{tabular}%
  \label{tab:cs5}%
\end{table}%
\begin{table}[htbp]
  \centering
    \fontsize{8}{10}\selectfont
  \caption{The rankings and scores predicted by the model for backward tail entity inference under varying information conditions are presented. The target entity is highlighted in bold.}
    \begin{tabular}{cccc}
    \toprule
    \multicolumn{3}{c}{George Clinton, artist record label music$^{-1}$, Casablanca Records} & Answer \\
\cmidrule{1-3}    test set & Top3 candidate entites & probabilities & Rank \\
    \midrule
    \multirow{3}[2]{*}{(h,r)} & Motown Records & 0.594  & \multirow{3}[2]{*}{7} \\
          & Jive Records & 0.554  &  \\
          & Atlantic Records & 0.531  &  \\
    \midrule
    \multirow{3}[2]{*}{(h,r)+(h,rs)} & Motown Records & 1.127  & \multirow{3}[2]{*}{6} \\
          & Jive Records & 1.035  &  \\
          & MCA Records & 1.006  &  \\
    \midrule
    \multirow{3}[2]{*}{(h,r)+(h,rs)+(p)} & \textbf{Casablanca Records} & 1.604  & \multirow{3}[2]{*}{1} \\
          & Motown Records & 1.481  &  \\
          & Jive Records & 1.389  &  \\
    \bottomrule
    \end{tabular}%
  \label{tab:cs6}%
\end{table}%

\end{document}